\documentstyle[11pt,newpasp,twoside,epsf]{article}

\begin{document}

\title{Dynamics of HCGs}

\author{H. M. Tovmassian, V. H. Chavushyan, O. Martinez, O. Yam}
\affil{Instituto Nacional de Astrof\'{\i}sica \'Optica y Electr\'onica, AP
51 y 216, 72000, Puebla, Pue, M\'exico}

and

\author{H. Tiersch}
\affil{Sternwarte K\"{o}nigsleiten, 81477, M\"{u}nchen, Leimbachstr.\ 1a 
Germany}

\begin{abstract}

It is shown that the radial velocity dispersion of HCGs is correlated with the
elongation of groups, $b/a$. It means that galaxies in a group move
preferentially along its elongation. Inspection of radial velocities of member
galaxies in chain-like and in round HCGs shows that galaxies in HCGs most
probably rotate around the gravitational center of the corresponding group.
Other two possible mechanisms: flying apart of galaxies from the group in
opposite directions, and infall of field galaxies upon the group are excluded.
It follows that HCGs are more stable formations, than it has been assumed, and
that the known inconsistencies between the results of N-body simulations and
the observational facts are not relevant.

\end{abstract}

\section{Introduction}

Hickson compact groups (HCGs) have been extensively studied during last
decades. Much of the interest raised because N-body simulations have shown
that members of HCGs should merge and form a single galaxy in a few crossing
times (Barnes 1985, 1989; Ishizawa 1986; Mamon 1987; Zheng, Valtonen, \&
Chernin 1993). More recent N-body simulations of poor groups (Bode et al.
1994; Athanassoula, Makino, \& Bosma 1997) indicated that supposed formation
of a central massive galaxy is a sufficiently rapid process. Meanwhile
observations do not prove the expected high rate of merging (Zepf 1993).
Moreover, it had been mentioned a strange absence of strong radio sources,
absence of strong signs of interaction, such as blue colors or FIR-emission,
the fact that the elliptical members (best candidates for ongoing mergers) are
not more frequently first-ranked than spirals, etc. (Menon 1995; Leon, Combes,
\& Menon 1998; Zepf 1993; Moles et al. 1994; Fasano \& Bettoni 1994; Pildis,
Bregman \& Schombert 1995; Sulentic 1997). To overcome the contradictions with
observations Mamon (1986, 1995), suggested that HCGs do not exist as physical
entities, and are a result of projection of galaxies in loose groups (LGs). On
the other hand, Hernquist, Katz \& Weinberg (1995), Ostriker, Lubin \&
Hernquist (1995) suggested that HCGs are filaments seen end-on. Contrary to
this there have been presented many evidences in favor of the physical reality
of HCGs (Hickson 1997 and references therein). To conciliate the results of
N-body simulations with the facts on the physical reality of HCGs, Governato,
Bhatia \& Chincarini (1991), Diaferio, Geller \& Ramella (1994, 1995)
suggested that CGs are being formed continually in collapsing LGs. This
mechanism, however, does not eliminate the mentioned contradictions.

\section{Regularity of movement of member galaxies in HCGs}

A clue to the solution of the problem was given by Tovmassian, Martinez, \&
Tiersch (1998). They showed that the radial velocity dispersion, $\sigma_{v}$,
of HCGs correlates with the elongation $b/a$\footnote{$a$ is the angular
distance between the most widely separated galaxies in the group, and $b$ is
the sum of the angular distances $b_{1}$ and $b_{2}$ of the most distant
galaxies on either side of the line $a$ joining the most separated galaxies
(Rood 1979).}. It was already known that HCGs have a triaxial spheroid
configuration with the true mean sphericity $\epsilon=0.26$ (Malykh \& Orlov
1986). Elongated groups with small $b/a$ have, on average, smaller
$\sigma_{v}$s, while more round groups with high $b/a$ ratio have higher
$\sigma_{v}$s. This finding showed that HCGs are real physical formations,
and that member galaxies in each HCG are moving along its elongation.

If member galaxies are indeed moving along the elongation of a group, then
$\sigma_{v}$s of chain-like groups with small two-dimensional galaxy-galaxy
median projected separation $R$ (supposed to be oriented at relatively small
angle to the line of sight) should, on average, be higher than that of groups
with higher $R$ (supposed to be oriented almost orthogonal to the line of
sight). There are 29 chain-like HCGs with $b/a\leq0.20$. The mean $R$ of 16 of
them with $R<40$ kpc is $26.3\pm7.0$ kpc, and the mean $\sigma_{v}$ is
$245.5\pm106.4$ km s$^{-1}$. The mean $R$ for 13 other groups with $R>40$ kpc 
is $68.4\pm26.0$ kpc, and the mean $\sigma_{v}$ is, as it was expected,
smaller, $137.5\pm134.3$ km s$^{-1}$. The Kolmogorov-Smirnov (K-S) test
showed that the hypothesis that both distributions are of the same parent
distribution is rejected with sufficiently high significance level
$\alpha=0.0065$. The groups of the considered two samples contain, on average,
practically the same number of member galaxies, 3.6 and 3.7. Thus, the found
correlation is not due to differences of the group masses. Hence, member
galaxies in HCGs are, indeed, moving preferentially along the elongation of
the corresponding group.

\section{Possible mechanisms of regular movement of member galaxies}

Regular movement of member galaxies preferentially along the main axis in
HCGs may be explained by the following three ways:

1. HCGs {\it disintegrate}, and their members {\it depart from each other}
in opposite directions.

2. Environmental galaxies {\it fall upon the group from opposite directions}
(Governato, Bhatia \& Chincarini 1991, Diaferio, Geller \& Ramella 1994, 1995).

3. HCGs are {\it gravitationally bound systems} the members of which {\it
rotate} around the gravitational center of the group in elongated orbits.

Disintegration of a CG may be explained by two possible mechanisms.

a. {\it Ejection} of galaxies in opposite directions from the central galaxy
according to Ambartsumian's (1961) or Arp's (1999, and references therein)
ideas. However, no any mechanisms of such processes are known, and we will
not consider this option.

b. {\it Disintegration} due to outer gravitational force. A group would
disintegrate if it is located between large clusters. However, the
distribution of orientation of the major axes of HCGs is isotropic in
relation to the large cluster-cluster alignment (Palumbo et al. 1993).
Hence, this option is not real. The gravitational influence of other CGs may
be another possibility. However, the velocity of galaxies in a group
initiated by the gravity force of another group with relatively small mass
would be much less than the observed values. Hence, there is no outside
acceptable force for stretching the groups. 

Occasional infall of environmental galaxies upon CGs may certainly take
place. Tovmassian, Yam, \& Tiersch (2001) and Tovmassian (2001) showed that
almost each HCG is associated with a LG members of which are distributed along
the elongation of the corresponding group. It was found that $\sigma_{v}$ of
members of a LG is correlated with the elongation of the whole HCG+LG system
(Tovmassian \& Chavushyan 2000). Hence, members of LGs also move
preferentially along the elongation of the corresponding system. This fact
seems to be in favor of the infall mechanism. However, as cosmological N-body
simulations showed the filaments are of a wide range of scales (West 1994,
Eisenstain, Loeb, \& Turner 1997), meanwhile the width of HCG+LG systems are
less than 125 kpc (Tovmassian 2001). It means that infall generally should be
{\it isotropic}. It is not explainable then why CGs are found in very narrow
filaments only in which infall may take place from opposite directions.
Moreover, according to Palumbo et al. (1993) groups of galaxies are not
generally aligned with either their nearest neighbors or with Abell clusters.
On the other hand, it is widely assumed that filaments are connecting large
clusters. It means then, that galaxies in CGs move not along the filaments. It
is, therefore, apparent, that occasional infall may not account for a
systematic regular movement of member galaxies. 

We are left, then, with the third option - rotation. If member galaxies rotate
around the gravitational center of the group, then in groups observed end-on
the galaxies which move towards the observer should be located on one side of
the cross-section of the group, while the galaxies which are moving in opposite
direction, should be located on the other side. In the case of infall of
galaxies upon the group there should be no any regularity in location of group
members. In six out of nine HCGs with $b/a\geq0.60$ (elongation of which is
oriented close to the line of sight) the galaxies which are moving towards or
from us are well separated on the cross-section of groups (Fig. 1). In only
two groups, HCG 20, and HCG 99, the galaxies with positive and negative
residual RVs are mixed over the cross-section of corresponding groups.

\begin{figure}
\plotone{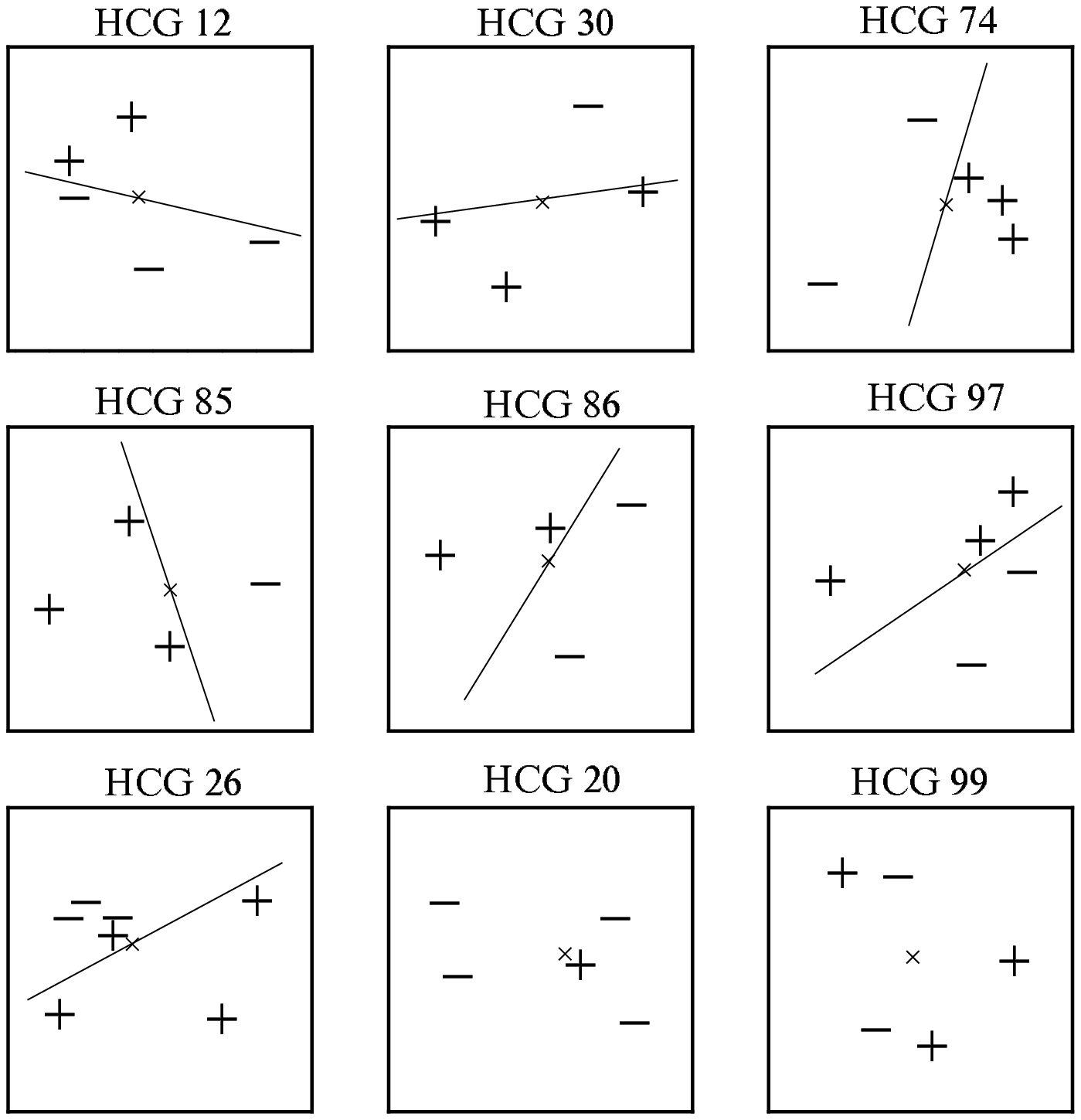}
\caption{The distribution of galaxies departing from us (+) and moving
towards us (-) over the cross-sections of round HCGs with $b/a>0.6$. Galaxies
with discordant redshifts are not shown.}

\end{figure}

Another test relates to chain-like groups seen side-on. If galaxies from
filaments fall upon the group, then the residual RVs of members of a LG
located at one side of the center of the CG+LG system at relatively large
distances from it, and that of those located on the other side should have
opposite signs. Inspection of RVs of compact and poor groups studied by
Carvalho et al. (1997), and Zabludoff \& Mulchaey (1998) shows that there is
no any regularity in distribution of signs of residual RVs in relation to the
center of the group.

Thus, the relative widths of cross-sections of CGs and filaments, and the
radial velocity data on HCGs and associated with them LGs {\it contradict} to
suggestion on inward motion of member galaxies. Observational data are {\it
in favor of systematic rotation} of member galaxies around the gravitational
center of the group. 

Using the mean value of the crossing time for HCGs, $\tau_{c} = 4/\pi \ R/V$,
where $R$ is the mean value of the two-dimensional galaxy-galaxy separation,
and $V$ is the intrinsic three-dimensional velocity dispersion, we may
roughly estimate the rotation time. $R$ may be determined more correctly in
the chain-like groups, and $V$ - in the round groups. For 29 HCGs with
$b/a\leq0.2$ the mean $R=45$ kpc, and for 24 round groups with $b/a>0.5$ the
mean $V$ is $\sim460$ km s$^{-1}$ ($R$ and $V$ values are taken from Hickson
et al. 1992). The mean $\tau_{c}$ is $\sim10^{8}$ years. If a HCG member
makes one revolution during $2\tau_c$ time, then according to Kepler's third
law, a LG galaxy at a mean distance of 250 kpc from the center of the CG
(Tovmassian 2001) makes one revolution in $\sim3\cdot10^{9}$ years, which is
less than the Hubble time. It follows that the faint members may be
gravitationally bound with corresponding group, and that the HCG+LG systems
may be virialized.

Hence, the third option, i.e. rotation of member galaxies around the
gravitational center of corresponding HCG is {\it the most realistic
mechanism}. HCGs are, thus, more stable systems than it has been predicted by
N-body simulations in which regular movement of member galaxies were not
taken into account. 

The probability is high that galaxies during their rotation around the
gravitational center of a HCG which has prolate spheroid configuration with
relatively small cross-section, would pass close to each other in its central
region. Here interactions between them could certainly occur. However, since
the velocities of galaxies near the pericenter of the orbit are sufficiently
high, the interaction time is short, the efficiency of interactions would be
small, and the merging processes would be rare. This explains the lack of
mergers and the lack of strong radio and FIR sources in HCGs. However, the
process of sweeping out the gas from member galaxies may be very effective,
and spirals would be converted to elliptical or lenticular galaxies. Indeed,
the fraction of spiral galaxies in HCGs, $f_{s}=0.49$, is significantly less
than in the field (Hickson, Kindle, \& Huchra 1988).

\section{The X-ray emission of the compact groups}

It is known that in clusters of galaxies $L_{x} \propto \sigma_v^{4}$ (Solinger
\& Tucker 1972, Quintana \& Melnick 1982), but groups do have more shallow
slope. It has been generally assumed that the reason is the enhanced (by
one-two orders of magnitude) X-ray emission of groups. The excess X-ray
luminosity of the low-mass groups is explained by the "mixed emission" scenario
(Dell`Antonio et al. 1994) when the emission from the intragroup plasma is
contaminated by a superposition of diffuse X-ray sources corresponding to the
hot interstellar medium of the member galaxies.

Tovmassian, Yam \& Tiersch (2002) showed that the shallow slope of $L_{x}
\propto \sigma_v$ line for groups is due to underestimation of $\sigma_v$s.
The mean $b/a$ of ten RASSCALS (Mahdavi et al. 2000) at the utmost left of the
$L_x - \sigma_{v}$ graph is equal to $0.44\pm0.11$, and that of the groups at
the utmost right is $0.68\pm0.19$. The mean value of $b/a$ of four HCGs
(Ponman et al. 1996) located most remote to the left of the line
$L_x-\sigma_{v}^4$ is equal to $0.20\pm0.05$, while that of the rest 18 groups
is equal to $0.44\pm0.18$. The K-S test rejected with the significance levels
$\alpha=0.0069$ and $\alpha=0.0082$ the hyphotesis that in both cases the
compared distributions are of the same parent distribution. Hence, the groups
located to the left of the $L_x \propto \sigma_{v}^4$ line on the $L_x -
\sigma_{v}$ graph have the smallest $b/a$ ratios. They are oriented close to
the orthogonal to the line of sight. For this reason their measured
$\sigma_v$ are smaller of the corresponding real value. Location of such
groups on the left side of the $L_x \propto \sigma_{v}^{4}$ line creates the
impression of the more shallow slope of the $L_x - \sigma_{v}$ correlation.
Hence, {\it not} the X-ray luminosities of the CGs are higher than it is
expected by the corresponding law for clusters of galaxies, but the
$\sigma_{v}$s of them are {\it underestimated}. Thus, consideration of
geometrical effect solves the longstanding problem of the X-ray emission of
groups of galaxies.

\section{Conclusions}

It is shown that $\sigma_v$s of HCGs are correlated with their elongation
(Tovmassian, Martinez \& Tiersch 1999). It means that member galaxies in a
group move preferentially along its elongation. It is concluded that members
of a HCG most probably rotate in elongated orbits around its gravitational
center. Other two possible explanations: the departure of member galaxies in
opposite directions, and infall of LG galaxies upon the CG from only opposite
directions are ruled out (Tovmassian 2001, 2002).

The finding that HCGs contain other members distributed along the elongation
of the corresponding groups (Tovmassian 2001, 2002; Tovmassian, Yam, \&
Tiersch 2001), and that their $\sigma_v$, as that of the proper members of
HCGs, correlates with the ellipticity of the whole system (Tovmassian \&
Chavushyan 2000) shows that HCGs are the cores of gravitationally bound poor
galaxy groups. 

Hence, HCGs are more stable systems, than it has been assumed. They have
elongated space configuration, contain more members located at relatively
large distances from them. The member galaxies most probably rotate around the
gravitational center of the corresponding group. Elongated space configuration
of HCGs, and regular rotation of their members around the corresponding
gravitational center are in agreement with the observed presence of galaxies
with signs of interaction, and smaller relative content of spirals. The lack
of apparently merged galaxies is also explained.

It is shown that the X-ray luminosity of CGs is not enhanced and obeys the
$L_{x} \propto \sigma_v^{4}$ found for clusters of galaxies. The discrepancy
is due to geometrical effect.

All members of elongated groups orientation of which is close to the line of
sight are seen near to each other on sky, though the distances between them
in space may be high enough. Such systems may often be recognized as poor
groups or even poor clusters. If orientation of elongated group is nearly
orthogonal to the line of sight, the group would be detected as a CG in the
case if its bright members happen to be close to each other on sky during
regular rotation around the gravitational center of the group.

\end{document}